\documentstyle[eqsecnum,prd,aps]{revtex}
\begin{document}
\draft
\title{COORDINATE-FREE SOLUTIONS FOR COSMOLOGICAL SUPERSPACE}
\author{D. S. Salopek}
\vspace{6pc}
\address{
{}Department of Physics and Astronomy, University of British Columbia,
Vancouver, Canada V6T 1Z1}

\date{March 5, 1997}

\maketitle

\begin{abstract}
Hamilton-Jacobi theory for general relativity provides an elegant covariant
formulation of the gravitational field. A general `coordinate-free'
method of integrating the functional Hamilton-Jacobi equation for gravity
and matter is described.  This series approximation method represents
a large generalization of the spatial gradient expansion that had been
employed earlier. Additional solutions may be constructed using a nonlinear
superposition principle.  This formalism may be applied to problems in
cosmology.

\end{abstract}


\widetext

\section{Introduction}

General relativity was formulated to describe the gravitational field in a
manner which
was independent of any particular choice of reference frame.
It is invariant under `general coordinate transformations'. However, when one
actually
solves the  Einstein field equations, say in a cosmological setting, one
typically
invokes a particular choice of
coordinates. As an improvement to this situation, one would prefer to utilize a
`coordinate-free' method of computing solutions to Einstein gravity.
In fact, Hamilton-Jacobi (HJ) theory for general relativity provides
such a description of the gravitational field \cite{SS92}, \cite{PSS94},
\cite{SS93}.
In solving the HJ equation, one need not specify the choice of time
hypersurface nor the
spatial coordinates.

Coordinate-free descriptions have proven useful in many fields of theoretical
physics. Vector analysis for Euclidean space is a good example.
Although one normally associates coordinates
with vectors, it is possible to interpret a vector geometrically in a
coordinate free
way using a magnitude and a direction.
In addition, by using his bra and ket notation, Dirac \cite{DIRAC} was able to
formulate
quantum mechanics in a form which was independent of the choice of basis
states.
The bra could be interpreted as an abstract vector without referring to a
particular
basis.

It has been known since the early 1960's that the HJ equation for general
relativity does not refer to the lapse and shift parameters which
characterize the gauge freedom of gravity \cite{Peres}. Hence the HJ equation
is the
natural starting point for a coordinate-free analysis of gravity.
However, the HJ equation is a nonlinear, functional differential equation
which characterizes an ensemble of universes,
{\it superspace}. It was generally believed that
superspace was too complicated to solve in its entirety. Beginning with
Misner \cite{MISNER}, researchers were content to solve homogeneous
minisuperspace models
where one considered only a finite number of degrees of freedom for the
gravitational field.
These investigations have been proven to be quite fruitful in quantum cosmology
(see, {\it e.g.}, Louko and Ruback \cite{LOUKO}).
(For a discussion of future trends in quantum cosmology, consult Hartle
\cite{HARTLE}
as well as Barvinsky and Kamenshchik \cite{B93}.)
However, there are some questions that these models cannot address since they
do not include inhomogeneities. For such models, the choice of time
hypersurface is
degenerate: a hypersurface of uniform $\phi$ is the same as that of uniform
scalar factor
$a$. In order to obtain a deeper understanding for the nature of time in a
semiclassical
setting \cite{S95}, \cite{ERICE95}, it is essential to include the role of
inhomogeneities.
(However, there are many additional obstacles in addressing the question of
time in
a quantum  context \cite{UNRUH}, \cite{KUCHAR}, \cite{LAFLAMME}.)

It is somewhat surprising that one can actually obtain solutions to
semiclassical
superspace using a series approximation method. The prototype solution for
the Hamilton-Jacobi (HJ) equation for general relativity utilized a spatial
gradient approximation. It was developed by
Salopek and Stewart \cite{SS92} and advanced further by Parry, Salopek and
Stewart
\cite{PSS94}.  In effect, they demonstrated that one can decompose superspace
into a
discrete sum of minisuperspaces. Here, a general method will be given for
constructing (HJ) solutions which utilize various other series approximations.
These techniques may be applied profitably to cosmology and other areas of
astrophysics.

A Hamilton-Jacobi description of general relativity is attractive for
several other reasons:

{\it (1) Primitive formulation of quantum theory for the gravitational field}.
In the semiclassical approximation, the wavefunctional is given by a
phase factor, $\Psi \sim e^{i {\cal S}/ \hbar }$, where
Planck's constant $\hbar$ is assumed to be tiny. The generating functional
${\cal S}$
then satisfies the HJ equation. If ${\cal S}$ is real, then one is describing
classical phenomena. If ${\cal S}$ is complex, one may describe quantum
phenomena such as tunnelling or the initial wavefunction of the universe.
It is generally believed that the fluctuations for structure formation as well
as microwave
background fluctuations are generated during an inflationary epoch.  For such
models,
it is absolutely essential to quantize the gravitational field, including both
scalar
and tensor modes (see, for example, \cite{SS95}, \cite{S95}). The current
status of inflation after the detection of the cosmic microwave background
anisotropy is discussed in refs. \cite{ERICE94}, \cite{BRANDENBERGER}.

{\it (2) Systematic solution of constraint equations}. In superspace, the
momentum
constraint may be given a simple geometric interpretation and it is easy to
solve.
The energy constraint fully characterizes
the dynamics of the gravitational system, and it is difficult to integrate
directly.
As will be demonstrated in the present work, it will be solved using a series
approximation.

{\it (3) Nonlinear analysis of gravity}. In the quasi-nonlinear regime,
one may employ HJ theory to analyze the formation of cosmological pancakes
\cite{Zel} during the matter-dominated epoch of the Universe \cite{CPSS94},
\cite{SSC94}.

Spatial gradient expansions in general relativity have a long history
\cite{GRADIENT}.
Within a HJ context, a spatial gradient expansion has been
applied successfully to cosmology including
a detailed computation of microwave background fluctuations and galaxy-galaxy
correlations
function arising from inflation \cite{SS95}, \cite{S92}.
Soda {\it et al} \cite{SODA} have generalized the expansion to encompass
Brans-Dicke gravity, and Chiba \cite{CHIBA} has formulated the gradient
expansion
for n-dimensional Einstein gravity.
The HJ equation for long-wavelength fields has often been invoked in an
attempt to recover the inflation potential from cosmological observations
\cite{SBB89}, \cite{SB}, \cite{COPELAND93}.

In Sec. II, the HJ equation for a scalar field interacting with
gravity is presented along with the analogous equation for a dust field
interacting with gravity. I quickly review the spatial gradient expansion
of the generating functional in a way which is easy to generalize to other
situations.
In Sec. III, the HJ equation including a scalar field is solved using a Taylor
series
in the scalar field $\phi$. In Sec. IV, the analogous method is applied for
a dust field, which describes collisionless, pressureless matter. In Sec. V,
the {\it Superposition Principle for Hamilton-Jacobi theory} allows one to
construct
complicated solutions of the HJ from known solutions which depend on a
parameter.
Conclusions and a summary follow in Sec. VI.

(Units are chosen so that $c=8\pi G=8\pi / m_P^2= 1$.
The  sign conventions of Misner, Thorne and
Wheeler \cite{MTW} will be adopted throughout.)

\section{Hamilton-Jacobi equation for general relativity with matter}

In the present work, two situations will be considered:
(1) a scalar field interacting with gravity and  (2) a dust field
interacting with gravity.

\subsection{Scalar field interacting with gravity}

The action for a single scalar field interacting with Einstein gravity is
\begin{equation}
{\cal I }=\int d^4 x \sqrt{-g} \left( {1\over 2}{\hspace{-0.1cm}}\ ^{(4)}
{\hspace{-0.04cm}}R
- {1\over 2} g^{\mu\nu}\partial_{\mu}\phi
\partial_{\nu}\phi - V(\phi)
\right);
\label{phifullaction}
\end{equation}
$\ ^{(4)}R$ is the Ricci scalar of the space-time metric $g_{\mu\nu}$.
For simplicity, the scalar field potential is assumed to be
\begin{equation}
V(\phi)= {1\over 2 } m^2 \phi^2 + V_0
\end{equation}
which describes a massive scalar field with a cosmological constant term.
(In general the solution methods described in this paper will work
for all potentials which are regular at $\phi=0$.)

In the ADM formalism the line element is written as
\begin{equation}
ds^2=\left(-N^2+\gamma^{ij}N_iN_j\right)dt^2 + 2N_idt\,dx^i +
\gamma_{ij}dx^i\,dx^j\ ,
\label{ADMdecomp}
\end{equation}
where $N$ and $N_i$ are the lapse and shift functions respectively,
and $\gamma_{ij}$ is the 3-metric. (For a elegant review, consult
D'Eath \cite{DEATH}).
One can then rewrite the action in Hamiltonian form,
\begin{equation}
{\cal I}=\int d^4x\left(\pi^{\phi} \dot\phi +\pi^{ij}\dot\gamma_{ij}
-N{\cal H} -N^i{\cal H}_i\right).
\label{phiADMaction}
\end{equation}
where ${\cal H}$ and ${\cal H}_i$ are the energy and momentum densities,
respectively.
In the Hamilton-Jacobi formalism, one replaces the momenta by functional
derivatives
of the generating functional, ${\cal S}[\gamma_{ab}(x), \phi(x) ]$,
\begin{equation}
\pi^{ij}(x)={\delta{\cal S}\over \delta{\gamma_{ij}(x)}}\ , \qquad
\pi^{\phi}(x)={\delta{\cal S}\over \delta\phi(x)}\ ;
\label{phi.pi}
\end{equation}
the generating functional associates a complex number for each field
configuration $\phi(x)$ on a space-like hypersurface whose
geometry is described by the 3-metric $\gamma_{ij}(x)$.
The energy and momentum constraints are now given by:
\begin{mathletters}
\begin{eqnarray}
0={\cal H}(x)&&=\gamma^{-1/2}
\left[2\gamma_{ik}(x) \gamma_{jl}(x) - \gamma_{ij}(x)\gamma_{kl}(x)\right]
{\delta{\cal S}\over \delta\gamma_{ij}(x)}
{\delta{\cal S}\over \delta\gamma_{kl}(x)} + \nonumber\\
&& {1\over 2} \gamma^{-1/2}\left({\delta{\cal S}\over \delta\phi(x)}\right)^2
+\gamma^{1/2} V[\phi(x)]  \;
 -{1\over 2}\gamma^{1/2}R
+ {1\over 2} \gamma^{1/2}\gamma^{ij}\phi_{i}\phi_{j}  \,  ,
\label{HJES}
\end{eqnarray}
and
\begin{equation}
0={\cal H}_{i}(x)=
-2\left(\gamma_{ik}{\delta{\cal S}\over \delta\gamma_{kj}(x)}
\right)_{,j} +
{\delta{\cal S}\over\delta\gamma_{kl}(x)}\gamma_{kl,i} +
{\delta{\cal S}\over\delta\phi (x)} \phi_{,i}  \, .
\label{MCS}
\end{equation}
\end{mathletters}
The Hamilton-Jacobi equation (\ref{HJES}) governs the evolution of the
generating functional,
${\cal S}\equiv {\cal S}[\gamma_{ab}(x), \phi(x)]$, in
{\it superspace}. It is quadratic in the momenta of fields.
The momentum constraint is linear in momenta.  Higgs \cite{Peres}
showed that it legislates that the generating functional is
invariant under reparametrizations of the spatial coordinates (`spatial
gauge-invariance').
It may be solved, for example, by assuming that the generating functional is an
integral of some function of the curvature, say $\int d^3x \gamma^{1/2} f(R)$,
or some other combination of spatial invariants.

\subsection{Dust field interacting with gravity}

The case of dust interacting with gravity is of high interest in cosmology.
It is generally believed that most of the Universe consists of dark-matter
whose
dynamics may be described by a dust field which is pressureless and
collisionless.
Secondly, hypersurfaces of uniform $\chi$ in general relativity come closest
to describing Lorentz frames that proved so useful in
flat spacetime. In fact, a Lorentz frame may be considered
to be a properly synchronized collection of dust particles. It may be necessary
to
introduce dust to interpret a quantum theory of the gravitational field
\cite{QUANTUM}, \cite{KUCHAR}.

The action
\begin{equation}
{\cal I }=\int d^4 x \sqrt{-g} \left[ {1\over 2}{\hspace{-0.1cm}}\ ^{(4)}
{\hspace{-0.04cm}}R
- {1\over 2} n\left( g^{\mu\nu}\partial_{\mu}\chi
\partial_{\nu}\chi + 1 \right)
\right],
\label{chifullaction}
\end{equation}
for a dust field, $\chi$, interacting with gravity,
is similar to that of a scalar field. The new ingredient is
the rest number density $n\equiv n(t,x)$ which is a Lagrange multiplier that
ensures that the 4-velocity
\begin{equation}
U^\mu = -g^{\mu\nu}\chi_{,\nu}
\label{4velocity}
\end{equation}
satisfies $U^\mu U_\mu = -1$. Hence $\chi$ may be interpreted as a velocity
potential.
In Hamiltonian form, the action is
\begin{equation}
{\cal I}=\int d^4x\left(\pi^{\chi} \dot\chi +\pi^{ij}\dot\gamma_{ij}
-N{\cal H} -N^i{\cal H}_i\right).
\label{chiADMaction}
\end{equation}
where $\pi^{\chi}$  is the canonical momentum of the dust field.
Replacing the momenta by functional derivatives
of ${\cal S}$,
\begin{equation}
\pi^{ij}(x)={\delta{\cal S}\over \delta{\gamma_{ij}(x)}}\ , \qquad
\pi^{\chi}(x)={\delta{\cal S}\over \delta\chi(x)}\ ,
\label{chi.pi}
\end{equation}
the energy and momentum constraint equations become:
\begin{mathletters}
\label{Sconstraints}
\begin{eqnarray}
0= {\cal H}(x)=&&\gamma^{-1/2} {\delta{\cal S}\over \delta\gamma_{ij}(x)}
{\delta{\cal S}\over \delta\gamma_{kl}(x)}
\left[2\gamma_{il}(x) \gamma_{jk}(x)
- \gamma_{ij}(x)\gamma_{kl}(x)\right] \nonumber \\
&& + \sqrt{1 + \gamma^{ij}\chi_{,i}\chi_{,j}}\,
{\delta{\cal S}\over\delta\chi (x)}
-{1\over 2}\gamma^{1/2}R  \, , \label{HJED} \\
0= {\cal H}_{i}(x)=&&-2\left(\gamma_{ik}{\delta{\cal S}\over
\delta\gamma_{kj}(x)}
\right)_{,j} +
{\delta{\cal S}\over\delta\gamma_{kl}(x)}\gamma_{kl,i} +
{\delta{\cal S}\over\delta\chi (x)} \chi_{,i} \, .
\label{MCD}
\end{eqnarray}
\end{mathletters}
In contrast to the Hamiltonian for a scalar field (\ref{HJES}), the energy
constraint
(\ref{HJED}) is {\it linear} in the canonical momentum of the dust field.
In a quantum context, the Hamiltonian constraint for dust and
gravity is thus very similar to the Schrodinger equation (or its covariant
generalization, the Tomonoga-Schwinger equation \cite{Tomo.Schw}) that has
been so successful in flat space-time.

\subsection{Review of Spatial Gradient Expansion}

The spatial gradient approximation for the HJ equation (\ref{HJES})
with a scalar field will be quickly reviewed. The essential aspects
will highlighted in order to illustrate what must be done in other
situations.

One expands the generating functional
\begin{equation}
{\cal S}[\gamma_{ab}(x), \phi(x)] =
\sum_{n=0}^\infty {\cal S}^{(2n)} \quad {\rm (spatial \; gradient \;
expansion)}
\label{theexpansion}
\end{equation}
in a series of terms according to the number of spatial gradient terms
that they contain.
It is quite important that each term in the series expansion satisfy the
momentum
constraint,
\begin{equation}
0= {\cal H}^{(2n)}_{i}(x) \equiv -2\left(\gamma_{ik}{\delta{\cal S}^{(2n)}\over
\delta\gamma_{kj}(x)}
\right)_{,j} +
{\delta{\cal S}^{(2n)}\over\delta\gamma_{kl}(x)}\gamma_{kl,i} +
{\delta{\cal S}^{(2n)}\over\delta\phi (x)} \phi_{,i} \, .
\end{equation}
The zeroth order term
\begin{equation}
{\cal S}^{(0)}=-2\int d^3x\,\gamma^{1/2}\, H\left(\phi\right) \, ,
\label{zeroth}
\end{equation}
is the simplest such term that one can imagine; it
contains no spatial gradients where the function $H(\phi)$ satisfies
\begin{equation}
H^2=    {2\over3} \left( {\partial{H}\over\partial\phi} \right )^2  + {1\over
3} V(\phi)\ .
\label{Hequation}
\end{equation}
The volume element  $d^3x\,\gamma^{1/2}$ appearing in eq.(\ref{zeroth})
is obviously invariant under spatial
coordinate transformations. The second order term contains two spatial
gradients and is an integral over the 3-curvature and a term quadratic
in spatial derivatives of $\phi$:
\begin{equation}
{\cal S}^{(2)}=\int \gamma^{1/2} d^3x
\left(J(\phi)R + K(\phi) \, \gamma^{ab} \phi_{,a} \phi_{,b}
\right) \, ;
\label{second.ansatz}
\end{equation}
$J$ and $K$ are arbitrary functions of $\phi$ which are chosen to satisfy the
second
order HJ equation. The higher order terms proceed along similar lines:
{\it e.g.}, ${\cal S}^{(4)}$ consists of all invariant terms with four
spatial derivatives.

What precisely is the expansion parameter in the gradient expansion?
To what does the index $2n$ refer?
The index $2n$ is related to the conformal weight of the functional ${\cal
S}^{(2n)}$.
To clarify this point, it is useful to introduce a scaling factor, $s$.
If one rescales the 3-metric using the homogeneous conformal factor, s,
\begin{equation}
\gamma_{ab}(x) \rightarrow s^2 \gamma_{ab}(x)
\end{equation}
one finds that
\begin{equation}
{\cal S}^{(2n)}[s^2 \gamma_{ab}, \phi] \; = \; s^{(3-2n)} \;
{\cal S}^{(2n)}[\gamma_{ab}, \phi]  \; .
\end{equation}
Hence the gradient expansion is an expansion of the generating functional in
powers
of the scaling factor $s$:
\begin{equation}
{\cal S}[s^2 \gamma_{ab}, \phi ] = \sum_{n=0}^\infty s^{(3-2n)}
{\cal S}^{(2n)}[\gamma_{ab}, \phi ]
 \; ,
\end{equation}
where ${\cal S}^{(2n)}[\gamma_{ab}, \phi]$ are simply the coefficients of
$s^{(3-2n)}$.
At the end of the calculation, one sets the scaling factor to unity because it
is
just a counting parameter.

The interpretation of the spatial gradient expansion as an expansion in powers
of
the conformal weight may be trivially extended to other situations. For
example,
instead of rescaling the metric, one may choose to rescale the matter field,
$\phi(x) \rightarrow s \phi(x)$ (or $\chi(x) \rightarrow s \chi(x)$) and then
expand in powers of $s$. Such a simple adjustment leads to radically different
forms of the generating functional as will be demonstrated in Sec. III and Sec.
IV.
(In order to reduce the introduction of extraneous notation,  one in practice
expands in powers of the desired field, and
typically foregoes any mention of the scaling factor $s$.)

\section{TAYLOR SERIES EXPANSION IN THE SCALAR FIELD}

For the case of a scalar field interacting with gravity, we will now consider
a solution for the generating functional in the form of a power series of the
scalar field:
\begin{equation}
{\cal S}[\gamma_{ab}(x), \phi(x)] =
\sum_{m=0}^\infty {\cal S}^{(m)} \quad {\rm (series \;\;in\;\;} \phi).
\label{scalar.expand}
\end{equation}
The zeroth order term will be assumed to be independent of $\phi$,
but otherwise, it is an arbitrary functional of the 3-metric,
\begin{mathletters}
\begin{equation}
{\cal S}^{(0)} \equiv {\cal S}^{(0)}[\gamma_{ab}(x)]  \; ,
\end{equation}
which is invariant under reparametrizations of the spatial coordinates:
\begin{equation}
0= -2\left(\gamma_{ik}{\delta{\cal S}^{(0)}\over
\delta\gamma_{kj}(x)}
\right)_{,j} +
{\delta{\cal S}^{(0)}\over\delta\gamma_{kl}(x)}\gamma_{kl,i}  \; .
\end{equation}
\end{mathletters}

\subsection{Equations for Scalar Field}

One substitutes the series into the HJ equation \ref{HJES}, and collects
terms of like order, to find:
\label{phi.equn}
\begin{mathletters}
\begin{eqnarray}
\left( {\delta{\cal S}^{(1)} \over\delta\phi(x)} \right )^2 =&&
 \gamma \, R - 2 \gamma \, V_0 +
- 2 \left[2\gamma_{il}(x) \gamma_{jk}(x)
- \gamma_{ij}(x)\gamma_{kl}(x)\right] \; {\delta{\cal S}^{(0)}\over
\delta\gamma_{ij}(x)}
{\delta{\cal S}^{(0)} \over \delta\gamma_{kl}(x)}\, ,
\quad {\rm (zeroth \; order \; terms)}\, ,  \label{phi1} \\
{\delta{\cal S}^{(1)} \over\delta\phi(x)} \,
{\delta{\cal S}^{(2)} \over\delta\phi(x)} =&&
-2  \left[2\gamma_{il}(x) \gamma_{jk}(x)
- \gamma_{ij}(x)\gamma_{kl}(x)\right] \; {\delta{\cal S}^{(0)}\over
\delta\gamma_{ij}(x)}
{\delta{\cal S}^{(1)} \over \delta\gamma_{kl}(x)}\, ,
\quad {\rm (first \; order \; terms)} \,  \label{phi2} \\
{\delta{\cal S}^{(1)} \over\delta\phi(x)} \,
{\delta{\cal S}^{(3)} \over\delta\phi(x)} = &&-{1 \over 2 } \,
\left( {\delta{\cal S}^{(2)} \over\delta\phi(x)} \right )^2
- { \gamma \over 2} \, \,  m^2 \phi^2
-{\gamma\over 2} \gamma^{ab} \phi_{,a} \phi_{,b}
 \qquad \qquad {\rm (second \; order \; terms)}  \label{phi3} \\
&& -  \left[2\gamma_{il}(x) \gamma_{jk}(x)
- \gamma_{ij}(x)\gamma_{kl}(x)\right] \; \left (
{\delta{\cal S}^{(1)}\over \delta\gamma_{ij}(x)}
{\delta{\cal S}^{(1)} \over \delta\gamma_{kl}(x)} +
2 {\delta{\cal S}^{(0)}\over \delta\gamma_{ij}(x)}
{\delta{\cal S}^{(2)} \over \delta\gamma_{kl}(x)} \right )
\, ,  \nonumber
\end{eqnarray}
\end{mathletters}
It is straightforward to derive a general expression for higher order
terms in analogy with ref. \cite{PSS94}.

\subsection{Solutions for a Scalar Field}

The generating functional ${\cal S}^{(n)}$ for any order may be written
in terms of a recursion relation. The first few terms are:
\begin{mathletters}
\begin{eqnarray}
{\cal S}^{(0)} \equiv && {\cal S}^{(0)}[\gamma_{ab}(x)]  \; , \label{phis0} \\
{\cal S}^{(1)} =&& \int d^3x \, \gamma^{1/2} \, \phi(x) \, \left [  R - 2 V_0
- 2 \gamma^{-1} \left[2\gamma_{il}(x) \gamma_{jk}(x)
-  \gamma_{ij}(x)\gamma_{kl}(x)\right] \;
{\delta{\cal S}^{(0)}\over \delta\gamma_{ij}(x)}
{\delta{\cal S}^{(0)} \over \delta\gamma_{kl}(x)} \right ]^{1/2} \,
\label{phis1} \\
{\cal S}^{(2)} = && -\int d^3 x \, \phi(x)
\left[2\gamma_{il}(x) \gamma_{jk}(x)
- \gamma_{ij}(x)\gamma_{kl}(x)\right] \; {\delta{\cal S}^{(0)}\over
\delta\gamma_{ij}(x)}
{\delta{\cal S}^{(1)}\over \delta\gamma_{kl}(x)}
\Bigg /  {\delta{\cal S}^{(1)} \over\delta\phi(x)} \, , \label{phis2} \\
{\cal S}^{(3)} = &&{1\over3 } \int d^3 x \, \phi(x) \, \Bigg \{
-{1 \over 2 } \,  \left( {\delta{\cal S}^{(2)} \over\delta\phi(x)} \right )^2
- {\gamma \over 2} \,  m^2 \phi^2
- {\gamma \over 2} \, \gamma^{ab} \, \phi_{,a} \phi_{,b}     \label{phis3} \\
&&  - \left[2\gamma_{il}(x) \gamma_{jk}(x)
- \gamma_{ij}(x)\gamma_{kl}(x)\right] \; \left (
{\delta{\cal S}^{(1)}\over \delta\gamma_{ij}(x)}
{\delta{\cal S}^{(1)} \over \delta\gamma_{kl}(x)} +
2 {\delta{\cal S}^{(0)}\over \delta\gamma_{ij}(x)}
{\delta{\cal S}^{(2)} \over \delta\gamma_{kl}(x)} \right )
\Bigg \} \Bigg / {\delta{\cal S}^{(1)} \over\delta\phi(x)} \, .
\nonumber
\end{eqnarray}
\end{mathletters}

The validity of these these formulae will be demonstrated by deriving
${\cal S}^{(2)}$.
The functional equation (\ref{phi2}) defining  ${\cal S}^{(2)}$ may be
rewritten in the form
\begin{equation}
{\delta{\cal S}^{(2)} \over\delta\phi(x)} =
-2 \left[2\gamma_{il}(x) \gamma_{jk}(x)
- \gamma_{ij}(x)\gamma_{kl}(x)\right] \; {\delta{\cal S}^{(0)}\over
\delta\gamma_{ij}(x)}
{\delta{\cal S}^{(1)} \over \delta\gamma_{kl}(x)} /
{\delta{\cal S}^{(1)} \over\delta\phi(x)}
\, . \label{temp1}
\end{equation}
Given ${\cal S}^{(1)}$, the right hand side is known. Hence eq.(\ref{temp1})
has the form of an infinite dimensional gradient which may be integrated using
a contour integral \cite{PSS94}
in $\phi$ field-space (the 3-metric is held fixed during such an integration).
For the same reasons given in earlier work \cite{S95},
one may choose an arbitrary contour of integration,
and I will use a straight line parameterized by
\begin{equation}
\overline \phi(x) = t \phi(x) \, , \quad 0 \le t \le 1 \, ,
\end{equation}
to connect the origin, $\phi_0(x) =0$ to the `final point' $\phi_f(x) =
\phi(x)$:
\begin{equation}
{\cal S}^{(2)} = \int d^3 x \, \int_0^1 dt \, \phi(x)
{\delta{\cal S}^{(2)} \over\delta \overline \phi(x)}
\end{equation}
After counting powers of $t$, the integral over $t$ gives $1/2$, and
hence ${\cal S}^{(2)}$ is given by eq.(\ref{phis2}).
Integrability of ${\cal S}^{(2)}$
in eq.(\ref{temp1}) is guaranteed if the previous order terms,
${\cal S}^{(0)}$ and ${\cal S}^{(1)}$,  are gauge-invariant \cite{S95}.

\subsubsection{First Example for a Scalar Field}

We will now give some explicit examples. Since ${\cal S}^{(0)}[\gamma_{ab}(x)]$
is arbitrary, the possibilities are limitless, but a sampling is instructive.

\begin{mathletters}
\begin{eqnarray}
{\cal S}^{(0)} = && - 2 H_0 \int d^3x \, \gamma^{1/2} \, , \quad H_0 =
\sqrt{V_0/3} \, , \\
{\cal S}^{(1)} = && \int d^3x \, \gamma^{1/2} \, \phi(x) \,  R^{1/2} \, , \\
{\cal S}^{(2)} = && -H_0 \, \int d^3x \, \gamma^{1/2} \,
\left [  \phi^2 + \left ( { \phi \over \sqrt{R} } \right )^{|k}
			  \left ( { \phi \over \sqrt{R} } \right )_{|k}
\right ] \, .
\end{eqnarray}
\end{mathletters}
No spatial derivatives of $\phi$ appear in ${\cal S}^{(1)}$. However,
they do appear in ${\cal S}^{(2)}$, and they cannot be removed by
integration by parts.

Unlike the spatial gradient expansion, the above series is not analytic since
${\cal S}^{(1)}$ contains a square root of $R$.
If $R<0$, ${\cal S}$ is imaginary, which is classically
forbidden. If the sign of the square root is chosen accordingly, these
solutions
would be exponentially suppressed in a quantum analysis:
\begin{equation}
\Psi \sim  e^{i {\cal S}} \, , \quad |\Psi|^2 = {\rm exp}
\left [- 2\int d^3x \gamma^{1/2} \, \phi \sqrt{-R}  \right ] \, , \quad
{\rm (for} \;\phi \ge 0).
\end{equation}

\subsubsection{Second Example for a Scalar Field}
Another simple example arises if we take ${\cal S}^{(0)} = 0$:
\begin{mathletters}
\begin{eqnarray}
{\cal S}^{(0)} = && 0 \, , \\
{\cal S}^{(1)} = && \int d^3x \gamma^{1/2} \, \phi \left( R - 2V_0 \right
)^{1/2} \, , \\
{\cal S}^{(2)} = && 0 \, , \\
{\cal S}^{(3)} = &&-{1\over 3} \int d^3 x \gamma^{1/2} \,  {\phi  \over
\sqrt{R - 2 V_0} } \,
\left [ { 1\over 2}  m^2 \phi^2 + {1\over 2 } \gamma^{ab} \phi_{,a} \phi_{,b} +
\gamma^{-1} \left[2\gamma_{il}(x) \gamma_{jk}(x)
- \gamma_{ij}(x)\gamma_{kl}(x)\right] \; {\delta{\cal S}^{(1)}\over
\delta\gamma_{ij}(x)}
{\delta{\cal S}^{(1)}\over \delta\gamma_{kl}(x)} \right ] \, ,
\end{eqnarray}

where
\begin{equation}
{\delta{\cal S}^{(1)} \over\delta\gamma_{ab}(x)} = {\gamma^{1/2}\over 2}
	\left [ (R-2V_0) \gamma^{ab} - R^{ab} + D^{|ab} - \gamma^{ab} D^2 \right ]
	\left [ \phi \left( R - 2V_0 \right )^{-1/2} \right ] \, .
\end{equation}
\end{mathletters}
If $R > 2 V_0$, then ${\cal S}^{(1)}$ describes the classically forbidden
regime.

\section{TAYLOR SERIES EXPANSION IN DUST FIELD}

\subsection{Equations for Dust Field}

I will now consider an expansion of ${\cal S}$ in powers of $\chi$:
\begin{equation}
{\cal S}[\gamma_{ab}(x), \chi(x)] =
\sum_{m=0}^\infty {\cal S}^{(m)} \quad {\rm (series \;\;in\;\;} \chi).
\label{dust.expand}
\end{equation}
This expansion is quite general, and it will be applicable in most instances
except
when one encounters a singular point (which will require a separate treatment).
The zeroth order term will be assumed to be independent of $\chi$:
\begin{equation}
{\cal S}^{(0)} \equiv {\cal S}^{(0)}[\gamma_{ab}(x)]  \; .
\end{equation}
Expanding all terms in powers of $\chi$, one obtains the following equations:
\begin{mathletters}
\label{chi.eq}
\begin{eqnarray}
{\delta{\cal S}^{(1)} \over\delta\chi(x)} = && {1 \over 2 } \gamma^{1/2} R
 -  \gamma^{-1/2} \left[2\gamma_{il}(x) \gamma_{jk}(x)
- \gamma_{ij}(x)\gamma_{kl}(x)\right] \;
{\delta{\cal S}^{(0)}\over \delta\gamma_{ij}(x)}
{\delta{\cal S}^{(0)} \over \delta\gamma_{kl}(x)} \, , \\
{\delta{\cal S}^{(2)} \over\delta\chi(x)} = &&
 -  2 \gamma^{-1/2} \left[2\gamma_{il}(x) \gamma_{jk}(x)
 -    \gamma_{ij}(x)\gamma_{kl}(x)\right] \;
{\delta{\cal S}^{(0)}\over \delta\gamma_{ij}(x)}
{\delta{\cal S}^{(1)} \over \delta\gamma_{kl}(x)} \, , \\
{\delta{\cal S}^{(3)} \over\delta\chi(x)} = &&
-{1 \over 2 }  \left( \chi_{|a} \chi^{|a} \right )
 {\delta{\cal S}^{(1)} \over\delta\chi(x)}
 -  \gamma^{-1/2} \left[2\gamma_{il}(x) \gamma_{jk}(x)
- \gamma_{ij}(x)\gamma_{kl}(x)\right] \; \left [
{\delta{\cal S}^{(1)}\over \delta\gamma_{ij}(x)}
{\delta{\cal S}^{(1)} \over \delta\gamma_{kl}(x)}  +
2 {\delta{\cal S}^{(0)}\over \delta\gamma_{ij}(x)}
{\delta{\cal S}^{(2)} \over \delta\gamma_{kl}(x)} \right ] \, .
\end{eqnarray}
\end{mathletters}

\subsection{Solutions for Dust Field}

The above equations (\ref{chi.eq}) may be integrated immediately:
\begin{mathletters}
\begin{eqnarray}
{\cal S}^{(0)} \equiv && {\cal S}^{(0)}[\gamma_{ab}(x)]  \, , \\
{\cal S}^{(1)} = && \int d^3x \, \chi(x) \, \left [ {1 \over 2 } \gamma^{1/2} R
 -  \gamma^{-1/2} \left[2\gamma_{il}(x) \gamma_{jk}(x)
- \gamma_{ij}(x)\gamma_{kl}(x)\right] \;
{\delta{\cal S}^{(0)}\over \delta\gamma_{ij}(x)}
{\delta{\cal S}^{(0)} \over \delta\gamma_{kl}(x)} \right ] \, , \\
{\cal S}^{(2)} = &&
 -  \int d^3 x \, \chi(x) \, \gamma^{-1/2} \left[2\gamma_{il}(x) \gamma_{jk}(x)
 -  \gamma_{ij}(x)\gamma_{kl}(x)\right] \;
{\delta{\cal S}^{(0)}\over \delta\gamma_{ij}(x)}
{\delta{\cal S}^{(1)} \over \delta\gamma_{kl}(x)} \, , \\
{\cal S}^{(3)} = && {1\over 3 }\int d^3 x \, \chi(x) \, \Bigg \{
-{1 \over 2 }  \left( \chi_{|a} \chi^{|a} \right )
 {\delta{\cal S}^{(1)} \over\delta\chi(x)} \\
 && -  \gamma^{-1/2} \left[2\gamma_{il}(x) \gamma_{jk}(x)
- \gamma_{ij}(x)\gamma_{kl}(x)\right] \; \left [
{\delta{\cal S}^{(1)}\over \delta\gamma_{ij}(x)}
{\delta{\cal S}^{(1)} \over \delta\gamma_{kl}(x)}  +
2 {\delta{\cal S}^{(0)}\over \delta\gamma_{ij}(x)}
{\delta{\cal S}^{(2)} \over \delta\gamma_{kl}(x)} \right ]  \Bigg \} \, .
\nonumber
\end{eqnarray}
\end{mathletters}
One may interpret the higher order terms, ${\cal S}^{(1)}$, ${\cal S}^{(2)}$,
$\ldots$,
as describing the evolution in time $\chi(x)$ of the the initial state
${\cal S}^{(0)}[\gamma_{ab}(x)]$.

\subsubsection{First Example for a Dust Field}

If ${\cal S}^{(0)}$ is proportional to the volume of a given 3-geometry,
the first few terms are:
\begin{mathletters}
\begin{eqnarray}
{\cal S}^{(0)} = && C \int d^3 x \gamma^{1/2} \, , \\
{\cal S}^{(1)} = &&  \int d^3 x \gamma^{1/2} \, \chi \,
\left [ {R \over 2 }  + { 3 C^2 \over 4 }  \right ] \, , \\
{\cal S}^{(2)} = && C \int d^3 x \gamma^{1/2} \left [
\chi^2 \left ( { R \over 8 }  + { 9C^2 \over 16 } \right )+
{1\over 2} \chi^{|k}  \chi_{|k} \right ]  \, .
\end{eqnarray}
\end{mathletters}

\subsubsection{Second Example for a Dust Field}

As a second example, consider a series whose first term is proportional
to an integral of the spatial curvature:
\begin{mathletters}
\begin{eqnarray}
{\cal S}^{(0)} = && E \int d^3 x \gamma^{1/2} \, R \, , \\
{\cal S}^{(1)} = &&  \int d^3 x \gamma^{1/2} \, \chi \,
\left [ {R \over 2 }  - 2E^2 \left( R^{ab} R_{ab} - {3 \over 8} R^2 \right )
\right ] \, .
\end{eqnarray}
\end{mathletters}

\section{SUPERPOSITION OF HAMILTON-JACOBI SOLUTIONS}

One of the very attractive features of quantum mechanics is the
principle of linear superposition: if $\psi_1 $ and $\psi_2$ are solutions of
the
quantum theory, then $\psi(y) = \psi_1 + \psi_2$ is a
solution as well.  This principle has no direct classical interpretation.
Nonetheless, there are situations where one can indeed construct additional
solutions
to the HJ equation from other known solutions.  One can enunciate a
{\it Principle of Superposition for Hamilton-Jacobi theory} which is motivated
by
a semiclassical treatment of the quantum theory.

For a quantum system with configuration variable, $y$,  suppose that
one is fortunate enough to find a solution to the Schrodinger equation,
$\psi(y|a)$ which depends on a {\it continuous} parameter $a$.
Then any linear superposition of these solutions is also a solution:
\begin{equation}
\psi(y) = \int da \, w(a) \,  \psi(y|a) \, , \label{quan.sup}
\end{equation}
where the weighting function $w(a)$ is arbitrary. If we work in the
semiclassical
limit, $\hbar \rightarrow 0$, then $\psi(y|a)$ and $w(a)$ may be approximated
by
phase factors,
\begin{equation}
\psi(y|a) \sim e^{i S(y|a)/ \hbar }, \quad w(a) = e^{i g(a) / \hbar} \, ;
\end{equation}
$S(y|a)$ is then a solution of the HJ equation which depends on a parameter
$a$. The resulting integral (\ref{quan.sup}) may approximated using the
stationary phase approximation,
\begin{mathletters}
\begin{equation}
\psi(y) = {\rm exp}[i ( S(y|a) + g(a) )/ \hbar ]
\end{equation}
where $a \equiv a(y)$ is now chosen so that the phase of the integrand has a
maximum or minimum,
\begin{equation}
0 = {\partial \over \partial a} \left [ S(y|a) + g(a) \right ]  \, .
\end{equation}
\end{mathletters}

Hence the principle of linear superposition in quantum mechanics and the
stationary phase approximation lead to
the {\it Principle of Superposition for Hamilton-Jacobi theory}:

\noindent
If  $S(y|a)$ is a solution of the HJ equation which depends on a continuous
parameter
$a$, then
\begin{mathletters}
\begin{equation}
T(y) = S(y|a) + g(a)
\end{equation}
is also a solution provided that $a \equiv a(y)$ is determined by the
stationary
phase condition,
\begin{equation}
0 = {\partial S(y|a) \over \partial a} + {\partial g(a) \over \partial a} \, .
\label{stat.phase}
\end{equation}
\end{mathletters}

The proof given above was motivated by the quantum theory.
In a gravitational context, it is not clear that a consistent
quantum theory exists (at present).  However, one can verify the principle
totally
within a classical context by noting that
\begin{equation}
{\partial T(y) \over \partial y} = {\partial S(y|a) \over \partial y}\Bigg|_a +
{\partial S(y|a) \over \partial a} \, {\partial a \over \partial y} +
{\partial g(a) \over \partial a} \, {\partial a \over \partial y} \, .
\end{equation}
The last two terms vanish by virtue of the stationary phase condition
(\ref{stat.phase}), and thus a derivative of $S(y)$ with respect to
$y$ coincides with a derivative of $S(y|a)$ with respect to $y$ (holding
$a$ fixed):
\begin{equation}
{\partial T(y) \over \partial y} = {\partial S(y|a) \over \partial y}\Bigg|_a
\, .
\end{equation}
Since $S$ appears in the HJ equation only in terms of its derivatives with
respect to the configuration variables, the principle is justified.

\subsection{Applying the Superposition to the HJ Equation for Gravity}

The superposition principle leads to some rather exotic solutions of the
HJ equation for general relativity.
To illustrate the basic principles, we will consider a massless scalar
field, $m=0$ with vanishing cosmological constant, $V_0 = 0$, which has the
following
spatial gradient expansion solution for the HJ eq.(\ref{HJES}),
\begin{mathletters}
\begin{equation}
{\cal S}[\gamma_{ab}(x), \phi(x) ] = {\cal S}^{(0)} + {\cal S}^{(2)} + \ldots ,
\label{m.0.V.0}
\end{equation}
of which the first two terms in a spatial gradient expansion are
\begin{eqnarray}
{\cal S}^{(0)} = &&-2C \int d^3 x \, \gamma^{1/2} \, e^{\sqrt{3/2} \phi}  \, ,
\\
{\cal S}^{(2)} = && \int d^3 x f^{1/2}
\left [  {1 \over  C}  e^{-4 \phi/ \sqrt{6}}
\left ( {1\over 8} R^f + {1\over 6} f^{ab} \, \phi_{,a} \phi_{,b} \right )
 + E R^{f} \right ]  \, .
\end{eqnarray}
\end{mathletters}
$C$ and $E$   are homogeneous but arbitrary parameters.
The new 3-metric $f_{ab}$ is related to $\gamma_{ab}$ by a conformal
transformation:
\begin{equation}
f_{ab} = e^{2\phi/ \sqrt{6}}   \gamma_{ab} \, .
\end{equation}
$R^f$ is the Ricci Scalar associated with $f_{ab}$.

Consider a new solution ${\cal T}$ of the HJ equation given by the sum
\begin{mathletters}
\begin{equation}
{\cal T} = {\cal S} - {1\over 2 } C^2  \, ,
\end{equation}
where ${\cal T}$ satisfies the stationary phase condition with respect to $C$:
\begin{equation}
0 = {\partial {\cal T} \over \partial C} = {\partial {\cal S} \over \partial C}
- C \, .
\end{equation}
\end{mathletters}
Using a spatial gradient expansion, one can solve for $C$:
\begin{mathletters}
\begin{equation}
C = C_0
- {1 \over  C_0^2} \int d^3 x f^{1/2} \, e^{-4 \phi /\sqrt{6}} \,
\left ( {1\over 8 } R^f + {1\over 6} f^{ab} \phi_{,a} \phi_{,b} \right ) +
\ldots \,
\end{equation}
where the functional $C_0$ is given by
\begin{equation}
C_0 = -2 \int d^3 x \, \gamma^{1/2} \, e^{\sqrt{3/2} \phi}  \, .
\end{equation}
\end{mathletters}
${\cal T}$ admits the following spatial gradient expansion:
\begin{mathletters}
\begin{equation}
{\cal T} = {\cal T}^{(0)} + {\cal T}^{(2)} + \ldots \,  \label{texp}
\end{equation}
with
\begin{eqnarray}
{\cal T}^{(0)} = && 2 \left ( \int d^3 x \, \gamma^{1/2} \,
e^{\sqrt{3/2} \phi} \right )^2 \, , \\
{\cal T}^{(2)} = && E \, \int d^3 x \, f^{1/2} R^f
+ \left [ \int d^3 x \, f^{1/2} \, e^{-4 \phi /\sqrt{6} }
\left ( \, -{1\over 16} R^f - {1\over 12 } f^{ab} \phi_{,a} \phi_{,b}\right)
\right ]
\Bigg / \left ( \int d^3 x \, \gamma^{1/2} \, e^{\sqrt{3/2} \phi} \right )
\, . \nonumber \\
\end{eqnarray}
\end{mathletters}
One may verify directly that eq.(\ref{texp}) is solution of the HJ
eq.(\ref{HJES}).

Unlike the original gradient expansion solution (\ref{m.0.V.0}) for ${\cal S}$,
the new solution ${\cal T}$  is no longer an integral over local terms, but
contains products and quotients of local integrals.  Hence very complicated
solutions of the HJ equations may be generated using the nonlinear
superposition of local solutions.

\section{CONCLUSIONS}
A coordinate-free analysis of general relativity
possesses distinct advantages over traditional methods of
solving Einstein's equations. When solving the field equations,
one typically makes arbitrary gauge choices. In many situations,
the optimal choice of gauge is not very clear, and a poor
choice of gauge can complicate the analysis significantly. In the
coordinate-free
method expounded here, one may solve the functional HJ equation without making
any gauge choices. However, applications of the HJ equation to physical
problems
have previously been hampered by the lack of mathematical tools.
Some general and useful techniques have been developed which
will be applied later in cosmology. However, the functional approach
is a radically different way of describing gravity.
More work is required in order to develop a more intuitive understanding
of the method.

Although one can obtain exact general solutions for gravitational superspace
in two spacetime dimensions (see, for example, \cite{LOUIS-MARTINEZ}),
it is doubtful that one may construct such solutions
in four spacetime dimensions. One must resort to some approximation method.
The spatial gradient expansion was the prototype solution of the HJ equation
for gravity and matter. It is basically a Taylor series
expansion in the conformal weight factor of the 3-metric.  By expanding
the generating functional in terms of other fields, one
may construct numerous other solutions to the HJ equation for general
relativity. Explicit expansions in either a scalar field $\phi$ or
a dust field $\chi$ were demonstrated explicitly. These solutions describe
the evolution of some gauge-invariant initial state ${\cal
S}^{(0)}[\gamma_{ab}(x)]$
as a functional of the matter field. Integrability of these
solutions is guaranteed by spatial gauge-invariance.

Many of these solutions
depend on continuous parameters. One can construct additional solutions by
using
the {\it Superposition Principle for Hamilton-Jacobi Theory} which is motivated
by the
superposition principle in quantum mechanics in conjunction with the stationary
phase approximation. One can in effect construct complicated solutions by
integrating
over the continuous parameters.

\acknowledgments

The author thanks J. M. Stewart, W. Unruh, D. Page, J. Soda and
D. Louis-Martinez for useful discussions.
This work was supported by the Natural Sciences and Engineering Research
Council of Canada (NSERC).

\vfill\eject

\begin{references}

\bibitem{SS92}
{D.S. Salopek and J.M. Stewart,
{\it Class. Quantum Grav.} {\bf 9}, 1943 (1992). }

\bibitem{PSS94}
{J. Parry, D.S. Salopek and J.M. Stewart,
{\it Phys. Rev. D} {\bf 49}, 2872 (1994).}

\bibitem{SS93}
{D.S. Salopek and  J.M. Stewart, {\it Phys. Rev.} {\bf D47},
3235 (1993).}

\bibitem{DIRAC}{P.A.M. Dirac, {\it The Principles of Quantum Mechanics}, 4th
edition,
(Oxford University Press, New York, 1958). }

\bibitem{Peres}
{P.W. Higgs, {\it Phys. Rev. Lett.} {\bf 1}, 373 (1958);
A. Peres, {\it Nuovo Cim.} {\bf 26}, 53, (1962).}

\bibitem{MISNER}{C.W. Misner, {\it Minisuperspace}, in {\it Magic Without
Magic: John
Archibald Wheeler}, ed. J. Klauder (W.H. Freeman, San Francisco, 1972).}

\bibitem{LOUKO}{J. Louko and P. Ruback, {\it Class.Quant.Grav.} {\bf 8}, 91,
1991.}

\bibitem{HARTLE}{J. B. Hartle,  {\it Quantum Cosmology: Problems for the 21st
Century},
in `Physics 2001', Nishinomiya-Yukawa Memorial Symposium on Physics in the 21st
Century:
Celebrating the 60th Anniversary of the Yukawa Meson Theory, Nishinomiya,
Hyogo, Japan, 7-8 Nov 1996 (1997).}

\bibitem{B93}
{A.O. Barvinsky and A.Y. Kamenshchik, Phys. Lett. {\bf B332},
270 (1994).}

\bibitem{S95}
{D.S. Salopek, Phys. Rev. {\bf D52}, 5563 (1995);
{\it ibid}, in Proc. of 1995 Canadian Conference on General Relativity,
(Fields Institute Publication, 1996).}

\bibitem{ERICE95} {D.S. Salopek, Phys. Rev. {\bf D52}, 5563 (1995);
D.S. Salopek, in Proceedings of the International School of Astrophysics
``D. Chalonge'', Fourth Course: String Gravity and Physics
at the Planck Scale, Current Topics in Astrofundamental Physics,
Erice, Italy, September 8-19, 1995, ed. N. Sanchez and A. Zichichi,
p.409-430 (Kluwer Academic Publishers, 1996).}

\bibitem{UNRUH}
{W.G. Unruh and R.M. Wald, {\it Phys.Rev.} {\bf D40}, 2598 (1989).}

\bibitem{KUCHAR}
{K.V. Kucha\v r, in Proc. of 4th Canadian Conference on
GR and Relativistic Astrophys.,
May 16-18 1991, ed. G. Kunstatter {\it et al}, p.211
(World Scientific, 1992).}

\bibitem{LAFLAMME}
{S.W. Hawking, R. Laflamme and G.W. Lyons,
{\it Phys.Rev.} {\bf D47}, 5342 (1993).}

\bibitem{ERICE94}
{D.S. Salopek, in Proceedings of the International School of Astrophysics
``D. Chalonge'', Third Course: Current Topics in Astrofundamental Physics,
Erice, Italy, September 4-16, 1994, ed. N. Sanchez and A. Zichichi,
p.179-204 (Kluwer Academic Publishers, 1995).}

\bibitem{BRANDENBERGER}
{R.H. Brandenberger, in
{\it 15th Symposium on Theoretical Physics: Field Theoretical Methods
in Fundamental Physics}, Seoul, Korea, 22-28 Aug 1996, ed. by
Choonkyu Lee (Mineumsa Co. Ltd., Seoul, 1997).}

\bibitem{Zel}
{Y.B. Zel'dovich,  A \& A,  {\bf 5}, 84  (1970).}

\bibitem{CPSS94}
{K.M. Croudace, J. Parry, D.S. Salopek and J.M. Stewart,
{\it Ap. J.} {\bf 423}, 22 (1994).}

\bibitem{SSC94}
{D.S. Salopek, J.M. Stewart and K.M. Croudace,
{\it Mon. Not. Roy. Astr. Soc.} {\bf 271}, 1005 (1994).}

\bibitem{GRADIENT}
{Lifshitz, E.M. \& Khalatnikov, I.M., 1964,
Usp. Fiz. Nauk, 80, 391 [Sov. Phys. Usp.,  6, 495 (1964);
Landau, L. \& Lifshitz, E.M., 1975, The Classical Theory of
Fields, (fourth English edition) Pergamon (Oxford);
K. Tomita, Progress in Theoretical Physics {\bf 54}, 730 (1975).}

\bibitem{SS95}{D.S. Salopek and J.M. Stewart, {\it Phys. Rev. D} {\bf 51}, 517
(1995).}

\bibitem{S92}
{D.S. Salopek, {\it Phys. Rev. Lett.}, {\bf 69}, 3602 (1992);
Proc. of the International School of Astrophysics,
``D. Chalonge'', Third Course: Current Topics in Astrofundamental
Physics, Erice, Italy, September 4-16, 1994, ed. N. Sanchez
(World Scientific Publishers, 1995).}

\bibitem{SODA}{J. Soda,  H. Ishihara and O. Iguchi,
Progress in  Theoretical Physics {\bf 94}, 781 (1995).}

\bibitem{CHIBA}{T. Chiba,  preprint KUNS-1277 (1995).}

\bibitem{SBB89}
{D.S. Salopek, J.R. Bond and J.M. Bardeen, {\it Phys. Rev. D} {\bf 40},
1753 (1989).}

\bibitem{SB}
{D.S. Salopek and J.R. Bond, {\it Phys. Rev. D} {\bf 42}, 3936 (1990);
{\it ibid}, {\bf 43}, 1005 (1991).}

\bibitem{COPELAND93}
{E.J. Copeland, E.W. Kolb, A.R. Liddle and J.E. Lidsey,
{\it Phys. Rev. D} {\bf 48}, 2529 (1993).}

\bibitem{MTW}
{C. Misner, K.S. Thorne and J.A. Wheeler,
{\it Gravitation} (New York: Freeman, 1973).  }

\bibitem{DEATH}{P. D'Eath, {\it Supersymmetric Quantum Cosmology}
(Cambridge University Press, 1996).}

\bibitem{Tomo.Schw}
{S. Tomonaga, in {\it Quantum Electrodynamics}, ed. J. Schwinger
(Dover, New York, 1958); J. Schwinger, {\it ibid.}.}

\bibitem{QUANTUM}
{D.S. Salopek, Phys. Rev. {\bf D46}, 4373 (1992).}

\bibitem{LOUIS-MARTINEZ}{D. Louis-Martinez, J. Gegenberg and G. Kunstatter,
Phys. Lett. {\bf B321}, 193 (1994); R.B. Mann, {\it ibid.} {\bf 294}, 310
(1992); A.M. Polyakov, {\it ibid.} {\bf 103}, 207 (1981). }

\end{references}
\end{document}